# ESCAPE

Preparing Forecasting Systems for the Next generation of Supercomputers

# D4.6 Report on workflow analysis for specific LAM applications

Dissemination Level: Public

This project has received funding from the European Union's Horizon 2020 research and innovation programme under grant agreement No 67162

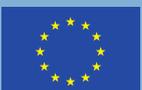 Funded by the European Union

Co-ordinated by

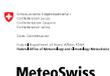 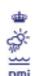 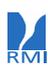 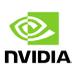 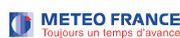 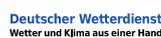 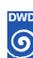 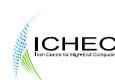 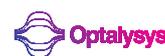 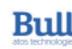 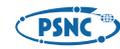 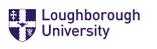

# Table of Contents

# 1 Executive Summary

In this deliverable we focus on the RMI-EPS ensemble prediction suite. We first provide a detailed report on the workflow of the suite in which 5 main categories of jobs are defined; pre-processing, lateral boundary conditions (LBCs), data assimilation, forecast and post-processing.

Combined Energy and wall-clock time measurements of the entire RMI-EPS suite were performed. They indicate that the wall-clock times are relatively spread between the various defined job categories, with the forecast accounting for the largest fraction at about 35%. As far as energy consumption is concerned, the forecast part dwarfs everything else and is responsible for up to 99% of the total energy consumption. This means that energy optimizations for the forecast part will translate almost proportionally into optimizations of the whole suite, while the maximum theoretical speed-up due to forecast optimizations cannot exceed a factor of about 3/2. Therefore, in terms of energy consumption, optimizations should first focus on the forecast part. For wall-clock time performance gains, however, optimizations (and possibly additional dwarfs) can be considered for the categories outside of the forecast part.

Finally, we report on our efforts to build a synthetic model of the suite through the `Kronos` workload simulator (cfr. The H2020 NEXTGenIO project). Such a synthetic model allows predicting the I/O and MPI behavior of the suite while subjected to hypothetical workloads on existing hardware. This is meant as a proof of concept and the necessary workflow is described without providing results of actual simulations.

# 2 Introduction

## 2.1 Background

ESCAPE stands for Energy-efficient Scalable Algorithms for Weather Prediction at Exascale. The project develops world-class, extreme-scale computing capabilities for European operational numerical weather prediction and future climate models. ESCAPE addresses the ETP4HPC Strategic Research Agenda 'Energy and resiliency' priority topic, promoting a holistic understanding of energy-efficiency for extreme-scale applications using heterogeneous architectures, accelerators and special compute units by:

- Defining and encapsulating the fundamental algorithmic building blocks underlying weather and climate computing;
- Combining cutting-edge research on algorithm development for use in extreme-scale, high-performance computing applications, minimizing time- and cost-to-solution;
- Synthesizing the complementary skills of leading weather forecasting consortia, university research, high-performance computing centers, and innovative hardware companies.

ESCAPE is funded by the European Commission's Horizon 2020 funding framework under the Future and Emerging Technologies - High-Performance Computing call for research and innovation actions issued in 2014.

### 2.2 Scope of this deliverable

#### 2.2.1 Objectives of this deliverable

The objectives can be summarized as follows:
- Use the performance simulators developed in Task 4.1 and the results of Task 4.4 (see D4.5) to make an external workflow analysis of the RMI-EPS ensemble suite.
- Document the current status of the suite.
- Provide an overall energy efficiency assessment of the suite. In contrast to D4.5, this includes the LBCs, data-assimilation as well as the post-processing stages.

#### 2.2.2 Work performed in this deliverable

A detailed description of the workflow of the RMI-EPS ensemble system is given. This description does not merely include the forecast stage but also what happens before and after the forecast, i.e. the boundary conditions (LBCs), the data assimilation and the post-processing stages.

An overall energy efficiency assessment of the suite is provided including all stages, i.e. LBCs, data assimilation, forecast and post-processing. Consequences of possible optimization strategies are discussed in relation to both energy consumption and wall-clock time.

We used the `Kronos` suite to build a synthetic model of the RMI-EPS suite. This synthetic model allows to predict the I/O and MPI behavior of RMI-EPS when subjected to hypothetical workloads on existing hardware. We report on this synthetic model as a proof of concept, without explicitly giving simulation results.

#### 2.2.3 Deviations and counter measures

Time constraints prevented the availability of the necessary `DCWorms` profiling of the RMI-EPS suite. As an alternative profiling system, we used the `Kronos` suite (cf. the H2020 NEXTGenIO project) to try and build a synthetic model of the RMI-EPS suite that allows monitoring its I/O and MPI behavior for hypothetical workloads on existing hardware.

## 3  RMI-EPS workflow

### 3.1  Setup

The RMI-EPS is a multi-model limited area ensemble prediction system consisting of 22 ensemble members, 11 using AROME physics and 11 using ALARO physics (Smet 2017). There are 2 control members, 1 using AROME physics and 1 using ALARO physics, and each control member has 10 corresponding perturbed members. While each ensemble member has a surface data assimilation cycle, only the two control members also possess a 3DVAR upper-air data assimilation cycle.

As already shortly described in Deliverable D4.3 (report on reference installations of several LAM models at ECMWF) a complete run of the RMI-EPS system consists of hundreds of inter-dependent jobs, roughly amounting to five stages: pre-processing, LBCs, data assimilation, forecast and post-processing, as shown in the RMI-EPS flowchart (Figure 1).

Pre-processing consists of checking whether the previous run has finished correctly, then fetching the IFS ensemble data and rearranging it in the correct data formats for the HarmonEPS system. This part runs on ECMWF's Linux cluster `ecgate` and could therefore not be included in the energy measurements and `Kronos` profiling. It should, however, only account for a small fraction of the total workload and energy consumption of the RMI-EPS system.

Subsequently, the LBCs, data assimilation, forecast, and (part of the) post-processing stages are executed on ECMWF's High Performance Computing Facility (`cca`). These are the stages that represent the bulk of the work in the RMI-EPS system, and for which energy consumption measurements and `Kronos` profiling are performed and described later in the document.

As a final post-processing step, some probability plots and other standard EPS products are created, archived and sent to the RMI (e.g. data interpolated to station locations). This final part again runs at ECMWF's Linux cluster `ecgate` and is therefore not included in the energy measurements and `Kronos` profiling, but its contribution is very minor compared to the bulk of the work performed at `cca`.

In the rest of this document we will describe in more detail the part of the RMI-EPS workflow that is executed on `cca`.

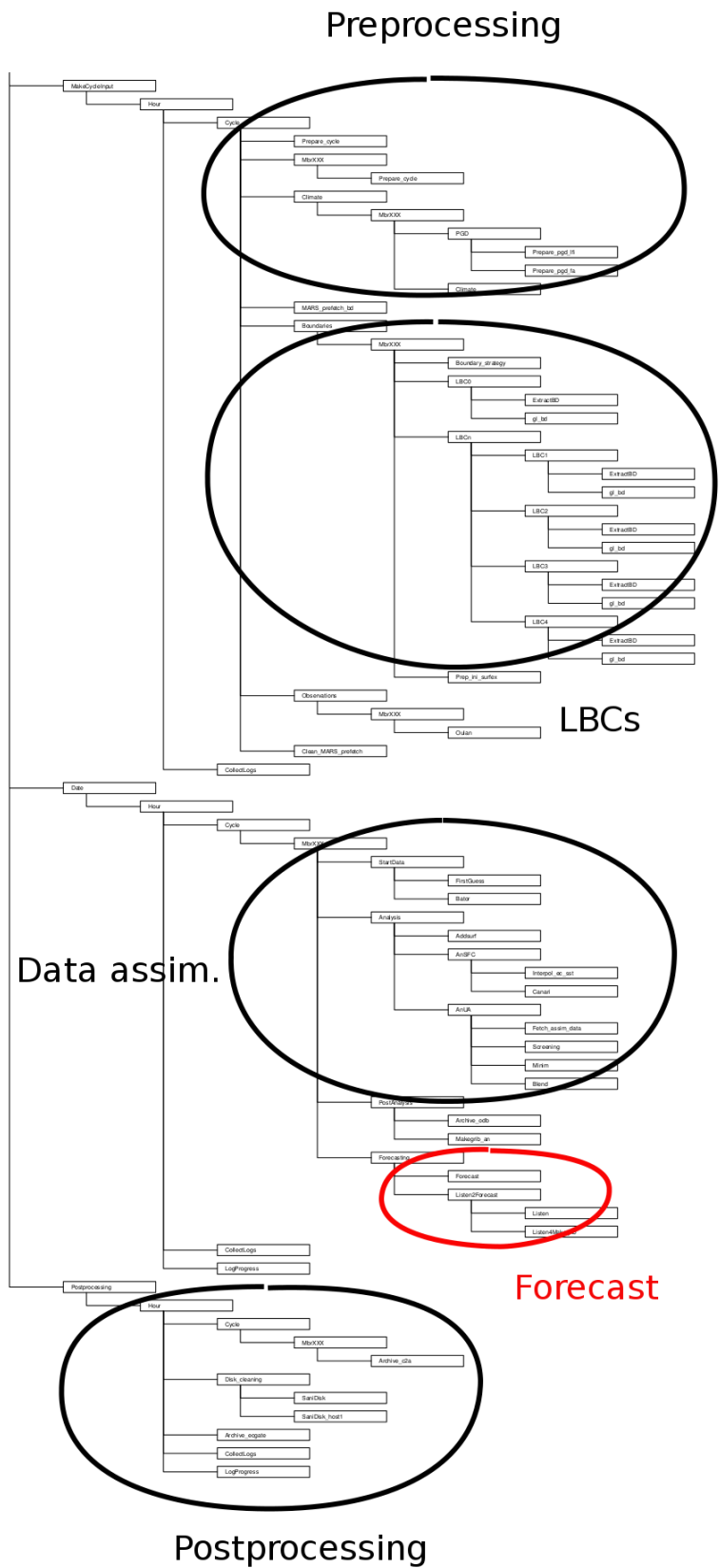

*Figure 1: RMI-EPS flowchart of individual jobs.*

### 3.2 Pre-processing and LBCs

The RMI-EPS workflow of the pre-processing and LBCs stage is shown in Figure 2. The jobs `MARS_prefetch_bd` and `ExtractBD` of the HarmonEPS system fetch boundary data from ECMWF's MARS archive but are not used in RMI-EPS. The necessary data is already fetched in the pre-processing stage on `ecgate` before HarmonEPS is started, so that these jobs become irrelevant for the most part. We will briefly describe the remaining jobs:

- `Prepare_cycle`: creates and cleans working directories.
- `Climate`: generates the climate files if necessary. This is only needed once a month, not for every run.
- `Boundary_strategy`: checks which boundary strategy should be used. In the case of RMI-EPS, the boundary data of ECMWF's EPS is used.
- `gl_bd` and `e927_bd`: preparation of initial and boundary files. Conversion from ECMWF's data files to ALADIN and AROME FA files, using either `gl` or `e927` (fullpos). Only one of these jobs has to be executed. In the case of RMI-EPS, the job `gl_bd` is used.
- `Prep_ini_surfex`: generates initial data for the SURFEX surface scheme using the FA files created above.

The job `gl_bd` has to be done for each lead time where the limited area model uses the global boundary data. In the case of RMI-EPS and a 36-hour lead time, the job `gl_bd` runs 13 times (per member), namely for the initial time, and for every 3 forecast hours. Also note that the jobs `gl_bd` and `Prep_ini_surfex` have to be executed for each of the 22 members.

### 3.3 Data assimilation and forecast

The RMI-EPS workflow for the data assimilation and forecast stages is shown in Figure 3. After the LBCs are created, and before the data assimilation of Figure 3, `Oulan` is run to extract conventional data and produce an OBSOUL file read by `BATOR`. This part can actually be seen as the first step of the data assimilation stage. Subsequently, the jobs in Figure 3 are submitted to `cca`. They can be briefly described as follows:

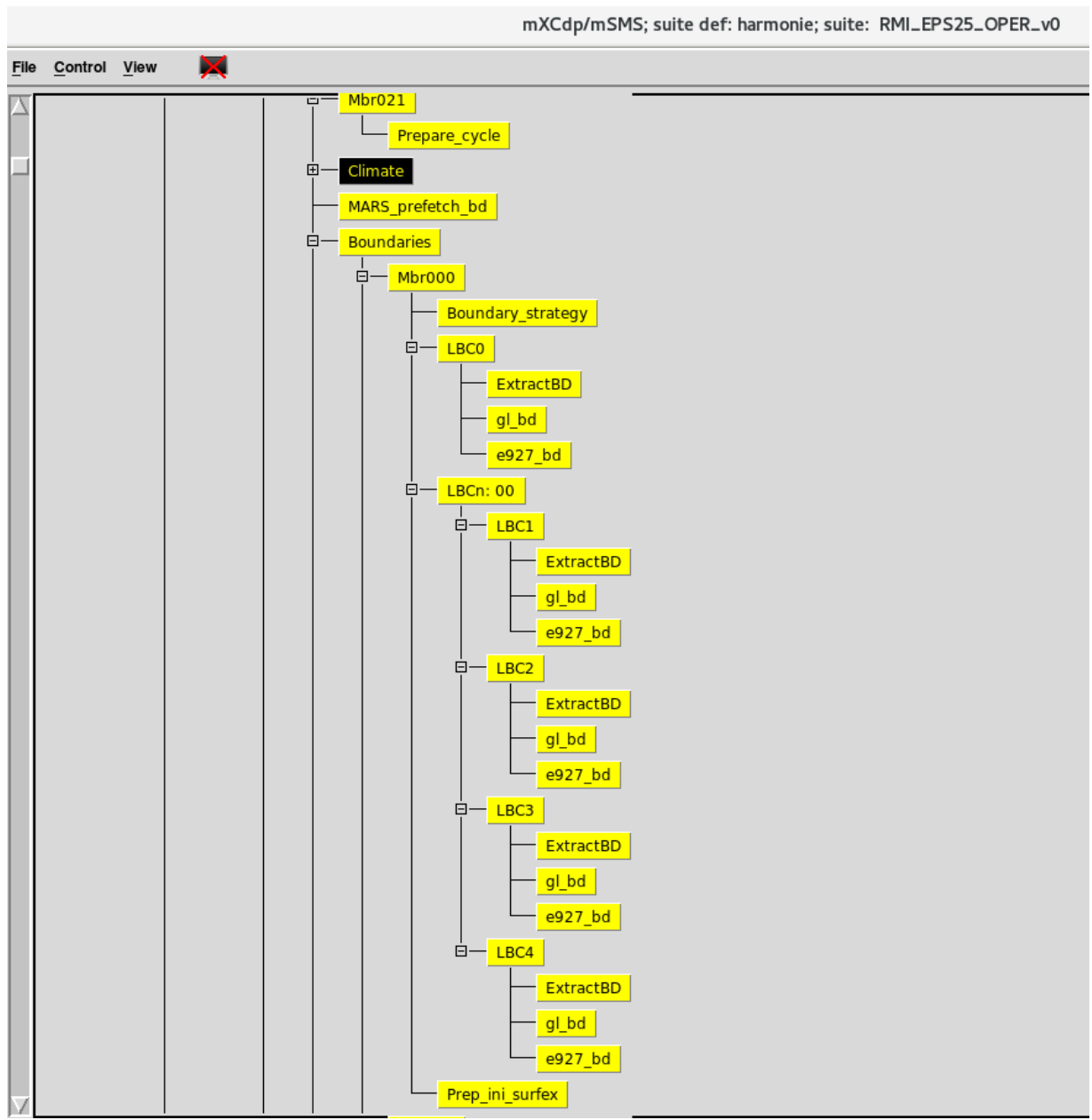

*Figure 2: RMI-EPS workflow: pre-processing and LBCs.*

- `FirstGuess`: extract first guess (analysis) for upper-air and SURFEX surface.
- `Bator`: preparation of ODB files used by data assimilation.
- `Addsurf`: add surface data to the first guess.
- `AnSFC_prep`: preparation of the surface data assimilation step.
- `Interpol_ec_sst`: interpolate sea surface temperature (sst) of ECMWF to model geometry of ALARO and AROME members.
- `Canari`: surface data assimilation using CANARI optimal interpolation method.
- `Fetch_assim_data`: preparation of the upper-air data assimilation step.
- `Screening`: quality control of observations to be used by the data assimilation.

- `Pertobs_ccma`: perturbation of observations. Not used in RMI-EPS system.
- `Minim`: minimization step of the upper-air 3DVAR data assimilation.
- `Blend`: blending step of the upper-air 3DVAR data assimilation.
- `Archive_odb`: archive `ODB` files.
- `Makegrib_an`: convert analysis `FA` files to grib files.
- `PertAna`: adding perturbations to the (upper-air) analysis of the control members to create (upper-air) analysis of the perturbed members.
- `Dfi`: Digital filter initialisation of the forecast. Not used in RMI-EPS system.
- `Forecast`: calculation of the forecast.
- `Listen2Forecast`: monitoring of the progress of the Forecast job.

Note that the above jobs usually have to be executed for each of the 22 ensemble members. The exceptions are the upper-air data assimilation jobs (`Fetch_assim_data`, `Screening`, `Minim`, `Blend`) which only have to be executed for the 2 control members, and the `PertAna` job which only has to be done for the 20 perturbed members.

### 3.4 Post-processing

The RMI-EPS workflow of the post-processing stage is shown in Figure 4. In fact, only the job `Archive_c2a` is really relevant for the RMI-EPS system. The jobs under `obsmonitor`, `Extract4ver`, `Fldver_family` and `field_monitor` are part of the HarmonEPS system to extract observations and forecasts at station locations for verification purposes, but these are not executed in the RMI-EPS implementation by default. Instead some additional post-processing is run at `ecgate`, as mentioned before. The job `Archive_c2a` saves all the initial and forecast data for all the 22 members to the ECFS archive of ECMWF.

### 3.5 Job categories for energy measurements

Above, we have described the bulk of the computation jobs needed for an RMI-EPS run. While so far, we used 5 stages to categorize the RMI-EPS jobs, it will nevertheless be more convenient to restrict ourselves to 4 main categories for the energy measurements (i.e. LBCs, data assimilation, forecast, post-processing) and a category "Others". This is also more correct, since some jobs in the data assimilation and forecast stage for instance, are strictly speaking neither part of the data assimilation nor the forecast (namely `Makegrib_an` and `PertAna`). We have chosen the job categories as described in Table 1:

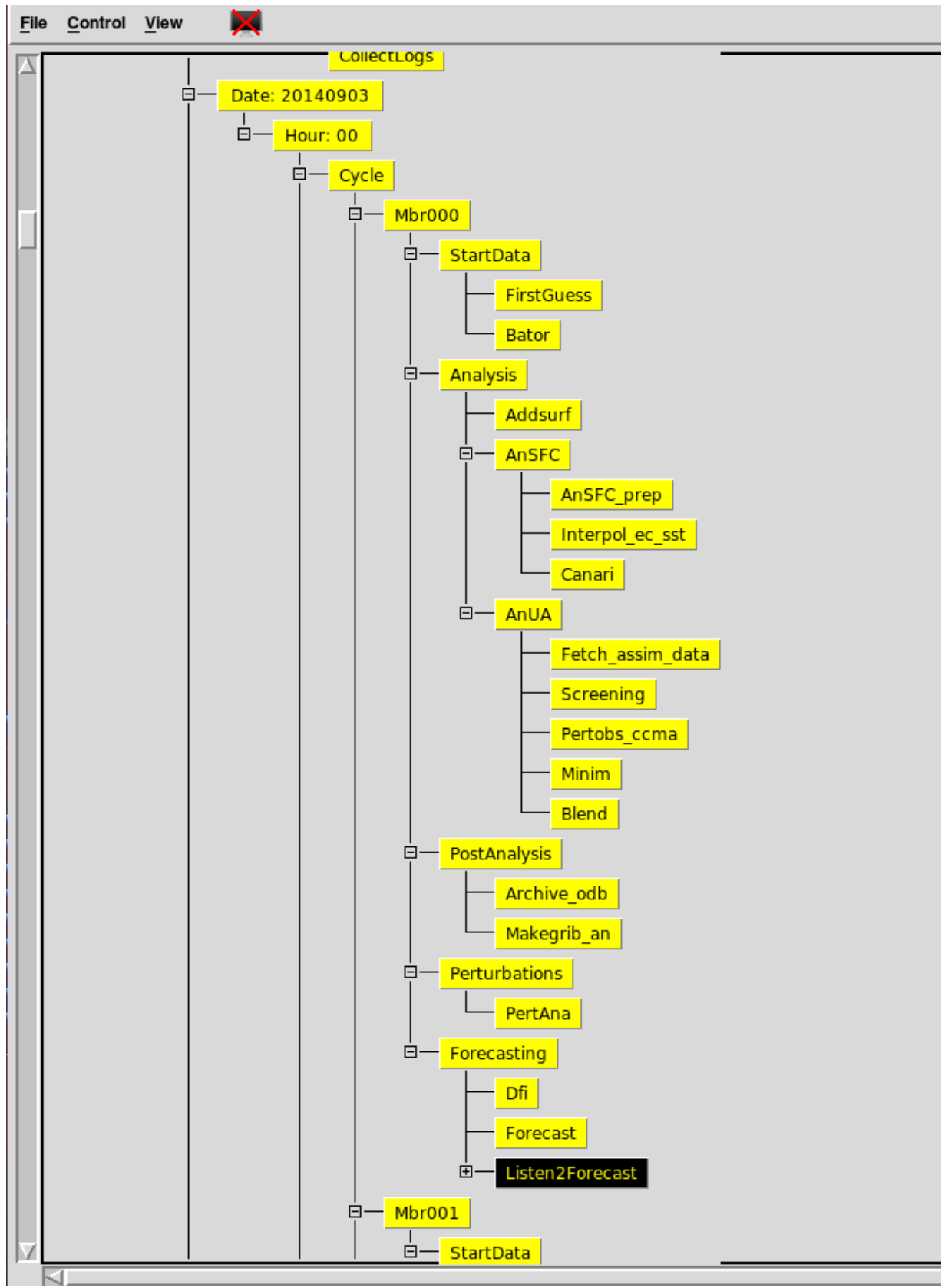

*Figure 3: RMI-EPS workflow: data assimilation and forecast.*

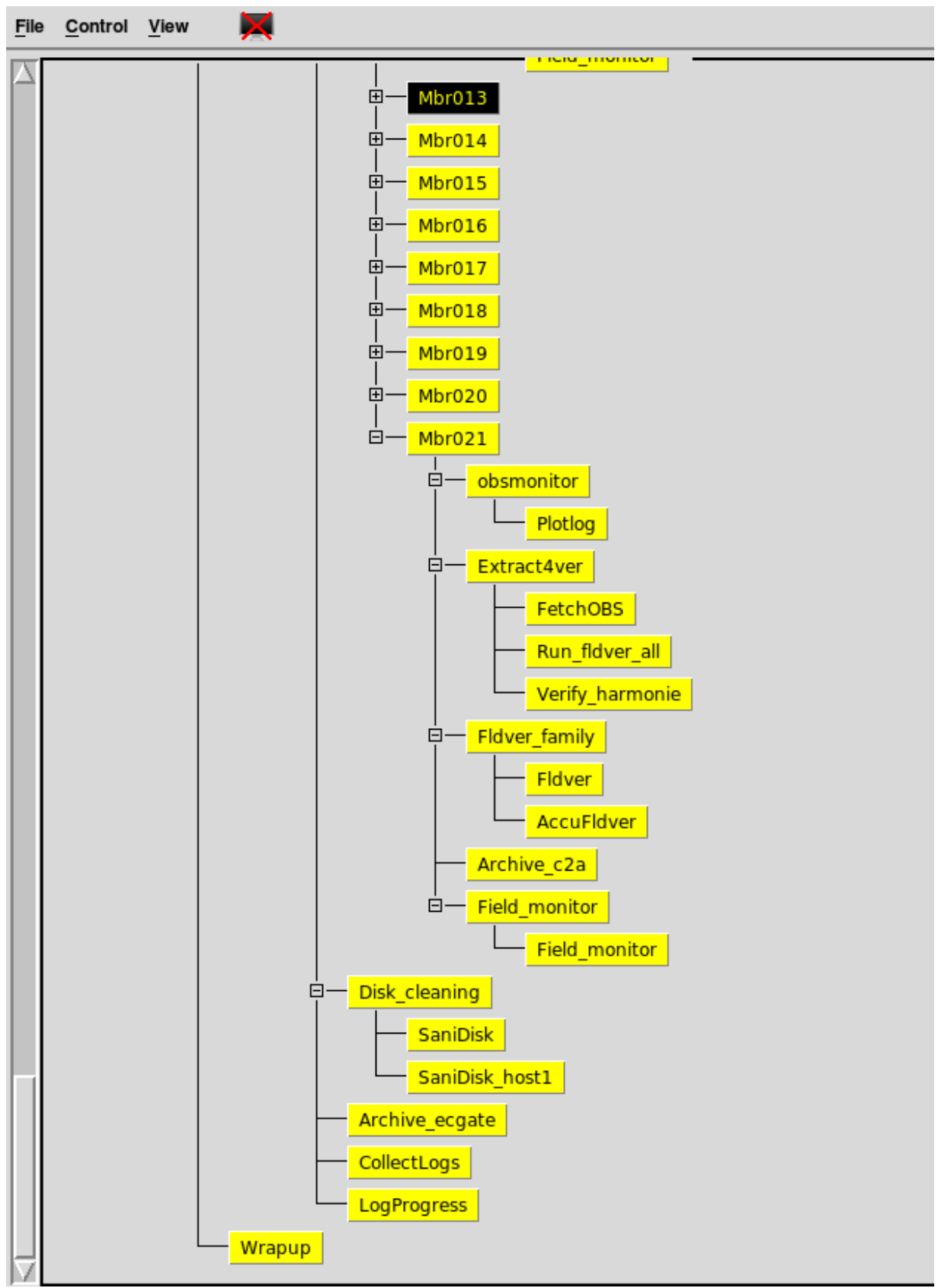

*Figure 4: RMI-EPS workflow: post-processing.*

| Stage | Job |
|---|---|
| LBCs | `MARS_prefetch_bd, Boundary_strategy, ExtractBD, gl_bd` |
| Data assimilation | `FirstGuess, Bator, Addsurf, AnSFC_prep, Interpol_efc_sst, Canari, Fetch_assim_data, Screening, Minim, Blend, Archive_odb` |
| Forecast | `Forecast, Listen2Forecast` |
| Post-processing | `Archive_c2a` |
| Other | `Prepare_cycle, Climate, Prep_ini_surfex, Makegrib_an, PertAna` |

*Table 1: categorization of the most important jobs within RMI-EPS suite.*

## 4   RMI-EPS energy measurements

On the `cca` cluster at the ECMWF, PAPI Cray Power Management (PM) counters provide access to the power management counters on the compute nodes. These counters enable the user to monitor and report energy consumed (in units of Joule) as well as elapsed time during program execution (in units of ms). The update frequency of these counters is about 10 Hz meaning that very short programs cannot be accurately measured. Furthermore, the counters do not allow the measurement of parts of a program (e.g. a subroutine or an individual loop).

On the basis of these counters, we implemented a script allowing for the simultaneous measurement of consumed energy as well as wall-clock time for any set of bash commands (e.g. an executable). The script in question needs to be called right before the said set of commands as well as right after. Note that the measurements only represent the consumption of energy due to the computational work of the compute nodes. We have no means of directly measuring any supplemental energy consumption like communication or cooling of the cluster.

`cca` is a Cray XC40 cluster, the measurements therefore pertain to the performance of the Intel Xeon E5-2695v4 "Broadwell" processors. The RMI-EPS setup consisted of a 36-hour forecast run including 2 control members and 20 perturbed members. Table 2 shows the resulting wall-clock time and energy measurements on a collection of the most demanding jobs. The used queue on `cca` and the used number of cores are also given. The jobs are categorized as in Table 1. We assume for simplicity, that a given job is run simultaneously for all the concerned ensemble members on separate nodes and so the contribution to the total wall-clock time is the average runtime of one such job (see also following paragraph for jobs that have a non-uniform workload per member). This does not hold for energy consumption, however, as runtime can be overlapped between members, but energy consumption must be added over all jobs and all members.

As mentioned in section 3.3, the workload of the control members for some jobs is not always identical to that of the perturbed members. Some jobs (like the upper-air data assimilation jobs `Fetch_assim_data`, `Screening`, `Minim` and `Blend`) are performed only by the 2 control members. At the other extreme, the `PertAna` job is performed only by the 20 perturbed members. For such jobs, we therefore decided to separate the energy consumption contributions of the number of control members (n) and the total number of members in the ensemble (N). This also allows one to extrapolate the results for other combinations of numbers of members. The wall-clock time per job is the average over all members in case of uniform workload, and the average of those members which have the highest workload (and therefore longest wall-clock time) in case of non-uniform workload.

Note that the job `gl_bd` is repeated 4 times per member; the first run is for the initial time (1 job), and subsequently 4 simultaneous jobs are launched once for every 3 forecast hours 9 (so 1+3*4=13, see sect. 3.2). Its entry in Table 2 is therefore computed as 4 times the wall-clock time of one job.

Jobs that have run on the `ns` queue on `cca` probably have energy contributions that are overestimated. This is because on that queue, the whole node is not reserved for the measured job and may therefore be contaminated by non-related jobs running on the same node. Their energy consumption entries are therefore orange colored in Table 1. Fortunately, their contributions are not significant.

| Stage | Job | Queue on `cca` | Ncores per member | Wall-clock time ctrl (sec) | Wall-clock time pert (sec) | Energy consumption (kJ) |
|---|---|---|---|---|---|---|
| LBCs | `MARS_prefetch_bd` | Ns | 1 | 12.3 | 12.3 | 1.7 |
| LBCs | `ExtractBD` | Ns | 1 | 2.6 | 2.6 | N * 0.4 |
| LBCs | `gl_bd` | Np | 36 | 266 | 266 | N * 98.5 |
| Data assim. | `FirstGuess` | Ns | 1 | 0.7 | 0.7 | N * 0.1 |
| Data assim. | `Bator` | Np | 36 | 335 | 181 | n * 20.7 + (N - n) * 11.6 |
| Data assim. | `Addsurf` | Np | 36 | 8.5 | 8.5 | N * 0.5 |
| Data assim. | `interpol_ec_sst` | Np | 36 | 11.7 | 11.7 | N * 0.8 |
| Data assim. | `Canari` | Np | 36 | 170 | 170 | N * 21.4 |
| Data assim. | `Screening` | Np | 324 | 212 | -- | n * 99.7 |
| Data assim. | `Minim` | Np | 324 | 82.3 | -- | n * 80.8 |
| Data assim. | `Blend` | Np | 36 | 11.0 | -- | n * 1 |
| Data assim. | `Archive_odb` | Ns | 1 | 121 | 7.9 | n * 21.2 + (N - n) * 1.1 |
| Forecast | `Forecast` | np | 612 | 1290 | 1290 | 0.5 * N * (4957.4+7982.3) |
| Post-proc. | `Archive_c2a` | np | 1 | 931 | 508 | n * 48 + (N - n) * 27 |
| Others | `Makegrib_an` | np | 36 | 82.1 | 36.0 | n * 7.6 + (N - n) * 3.1 |
| Others | `PertAna` | ns | 1 | -- | 31.6 | (N - n) * 5.3 |
| Total energy consumption | | | | | | ≈ 230.9 * n + 6639.6 * N + 1.7 |
| Total for n=2 and N=22 | | | | | | ≈ 146000 |

*Table 2: Wall-clock time and energy consumption of the most demanding jobs within the RMI-EPS ensemble suite. The jobs are listed according to the stage to which they belong within the workflow of the suite (see first column). The energy consumption results are written according to their dependency on the total number of ensemble members (N) and the number of control members (n) used (see text).The wall-clock times correspond to n=2 and N=22 which is the current setup of the RMI-EPS suite.*

Using the results of Table 2, pie charts for both the wall-clock times and the energy consumption contributions per job category for the control members resp. the perturbed members are shown in Figure 5 and 6. It is clear that, in terms of the energy consumption, the Forecast job dominates everything else in all situations, independently of the number of ensemble members or control members used. This can be explained by the relatively long wall-clock times as well as the large number of cores used for the job per member. This situation does not change as the total number of members is increased. The immediate consequence of this is that any future energy optimization gain obtained by the ESCAPE dwarfs on the Intel Broadwell processors will be transferred almost fully into the energy performance of the RMI-EPS ensemble suite as a whole.

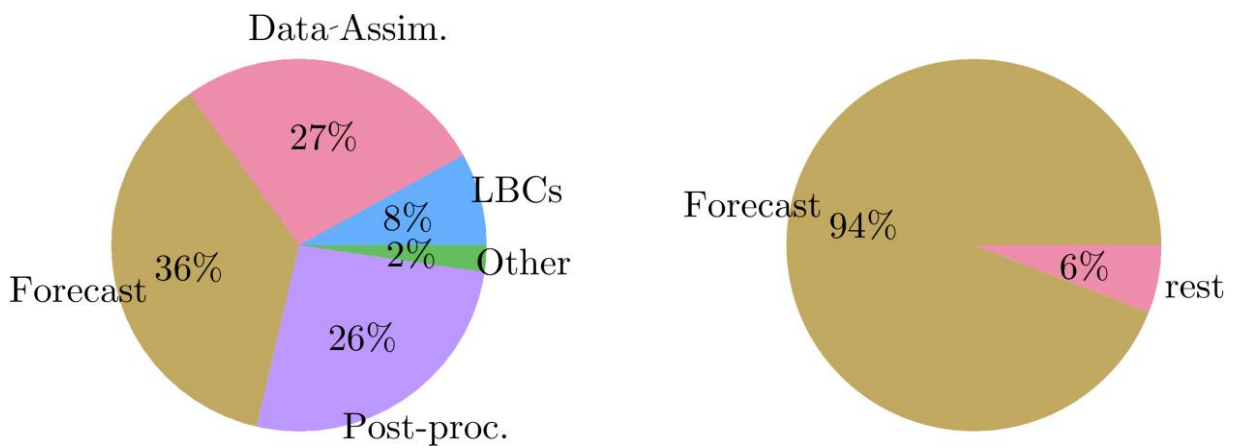

*Fig 5: Distribution of wall-clock times (left) and energy consumption (right) for the 5 job-categories as applied to the control members of the ensemble.*

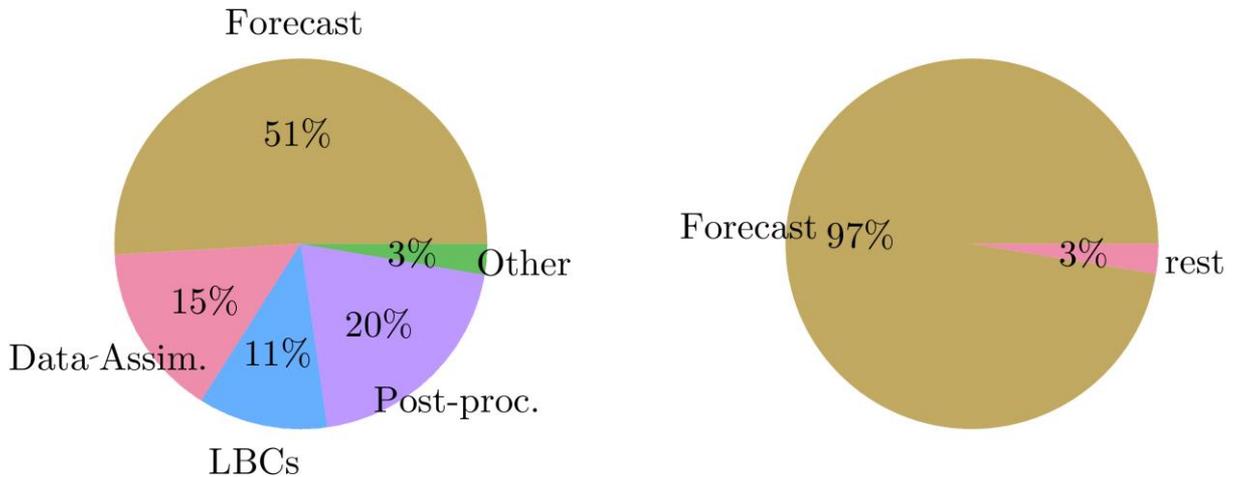

Fig 6: Distribution of wall-clock times (left) and energy consumption (right) for the 5 job-categories as applied to the perturbed members of the ensemble.

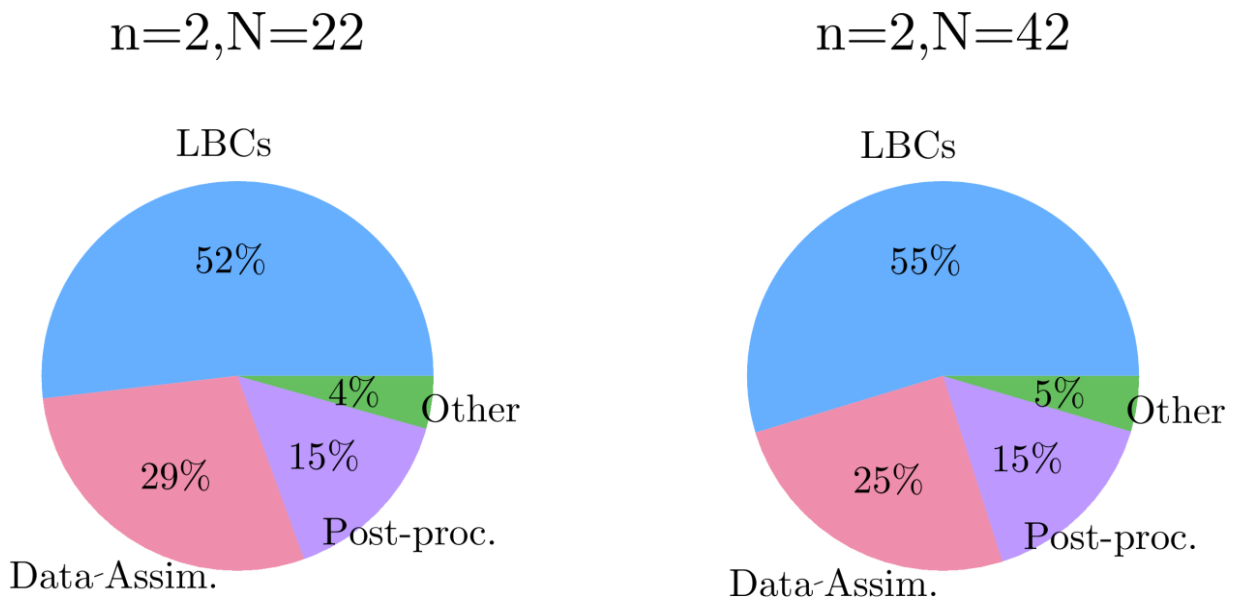

Fig 7: Distribution of energy consumption for the entire suite for 2 different number of perturbed ensemble members (20 for the left chart and 40 for the right chart). Here we restricted ourselves to the non-Forecast categories of the jobs.

In terms of wall-clock time, the relative contributions are more evenly distributed. The Forecast still remains the dominating job for both the control members and the perturbed members, but now the contributions of the other categories cannot be neglected, except perhaps for the 'Other' category. The differences between the workloads of the control members and perturbed members are clearly visible. Even though the forecast wall-clock times between both types of members is the same, it is the larger workload of the control members during the data-assimilation and post-processing stages which results in a larger forecast fraction for the perturbed members (51% versus 36%).

Figure 7 compares the distribution of energy consumption for the cases (n=2, N=22) and (n=2, N=42). Since the Forecast dominates the total energy consumption in both cases, we here focus on the relative contributions of the remaining job categories to the total consumption minus that of the Forecast. The main change occurs in the decrease of the contribution of the data assimilation category. This is expected however, since we increased the number of perturbed members combined with the fact that data assimilation is only performed by the control members.

Figure 8 plots of the energy consumption vs. wall-clock time for individual jobs for the case of (n=2, N=22). Added are lines of constant power which give an indication of the average power at which individual jobs are running. The Forecast job is clearly above everything else, consuming power at a rate of about 5kW (due to the large number of used cores, see Table 2). Also added as a red line is the measured power consumption of one node in an idle state (by measuring the energy consumption during a 1 minute sleep command). It is interesting to see that a few jobs lie in the vicinity of this line (depending of course on the accuracy of our measurement).

Summarizing the results, if future wall-clock performance optimizations only focus on the Forecast part, then the only gains achieved will be in the part of the code accounting for a maximum of about half of the wall-clock time. This means that in that case the expected maximum theoretically achievable speedup for the RMI-EPS suite would be about 2. It is thus advisable to look for new dwarfs in the categories outside of the Forecast to substantially improve the wall-clock time performance of large ensemble systems.

Fig 8: Energy consumption versus wall-clock time for a collection of individual jobs of one of the control members of the ensemble. The red line represents an estimate of the power consumption of one node during an idle state (see text).

## 5   Kronos

### 5.1   Introduction

`Kronos` is a benchmarking suite developed at the ECMWF as part of the NEXTGenIO project. `Kronos` differs from other benchmarking tools in that it is able to model all of the components of an HPC system simultaneously instead of just one of its components (e.g. I/O, compute, network and scheduling) (Bonanni et al. 2016). It reads previously obtained profiling information of the code to be modelled and uses that to generate synthetic workloads which can then be used to simulate (scaled versions of) the original code on the `same` HPC system/cluster. It cannot, however, model the behaviour of existing codes on future hardware systems nor is it designed to profile energy consumption. It therefore serves a different purpose compared to DCWorms.

`Kronos` can, however, be of use in the context of tenders by providing a synthetic model of the code that can more easily be run and benchmarked by different vendors. It is also able to model a scaled up version of the original workload. This would allow vendors to benchmark the code on their hardware using various scenarios without actually having to recompile the code or make different initial setups. In this section we attempt to demonstrate the viability of a synthetic `Kronos` workload model of the RMI-EPS suite.

### 5.2   Profiling tools

We profiled the RMI-EPS suite installed on `cca` at the ECMWF with the `Darshan` and `IPM` profiling tools. The choice of these 2 profiling tools was guided by the fact that the `Kronos` suite possesses native ingestion modules for them. However, the precise profiling tools used are not decisive for this proof of concept.

#### 5.2.1   IPM

IPM is a portable profiling infrastructure for parallel codes. It provides a low-overhead performance profile of the performance aspects and resource utilization in a parallel program. Communication, computation, and I/O are the primary focus. The two main objectives of IPM are ease-of-use and scalability in performance analysis. It does not require code modification.

To use IPM on HPC architectures that support shared libraries, all that is needed is to load the `ipm` module. Once the module is loaded, jobs can be run as normal and a performance profile is written once the job has successfully completed. Relinking the code is not required. For static executables and architectures which do not support shared libraries however, relinking is in fact required.

### 5.2.2 Darshan

Darshan is a scalable HPC I/O characterization tool. It is designed to capture an accurate picture of application I/O behavior, including properties such as patterns of access within files, with minimum overhead. Darshan is particularly suited for storage research in the context of HPC computing. As for IPM, Darshan requires no code modification in our case. Similarly, to IPM, one loads the appropriate `darshan` module to setup the correct environment and subsequently runs the code. The profiling results are then written to a prescribed location.

Darshan also possesses a 'single' mode which is able to profile non-parallel applications. This allows to profile every single bash command inside a script, for instance, which is necessary as I/O contributions also occur in those parts of the workflow which are not parallellized. One does need to switch to a `darshan-single` module for this mode to be active, however.

### 5.3 Assembling a synthetic model of RMI-EPS

For our purposes, we can summarize the `Kronos` modeling process as follows:

1) **Profiling:** the code to be modeled is profiled with the help of a set of profiling tools (in our case Darshan and IPM).
2) **Data ingestion**: the profiling results are turned into `Kronos`-specific data format files.
3) **Schedule generation**: schedules for the modeled code are built based on the dependencies of any individual sub-jobs and the results of step 2.
4) **Execution**: these abstract schedules are used by the `Kronos` Executor which submits concrete synthetic jobs to the scheduling queue of a real HPC system.

The complete RMI-EPS ensemble suite contains hundreds of individual jobs. Trying to adapt each script to enable profiling would be prohibitively time-consuming. We instead chose to focus on a set of the about 20 most demanding jobs in terms of wall-clock time. Figure 9 shows the profiled jobs and dependencies graphically. We made sure that the major part of the suite workflow was represented, i.e. we included the dominant jobs of the pre-processing, the data-assimilation, the Cycle (forecast) and the post-processing.

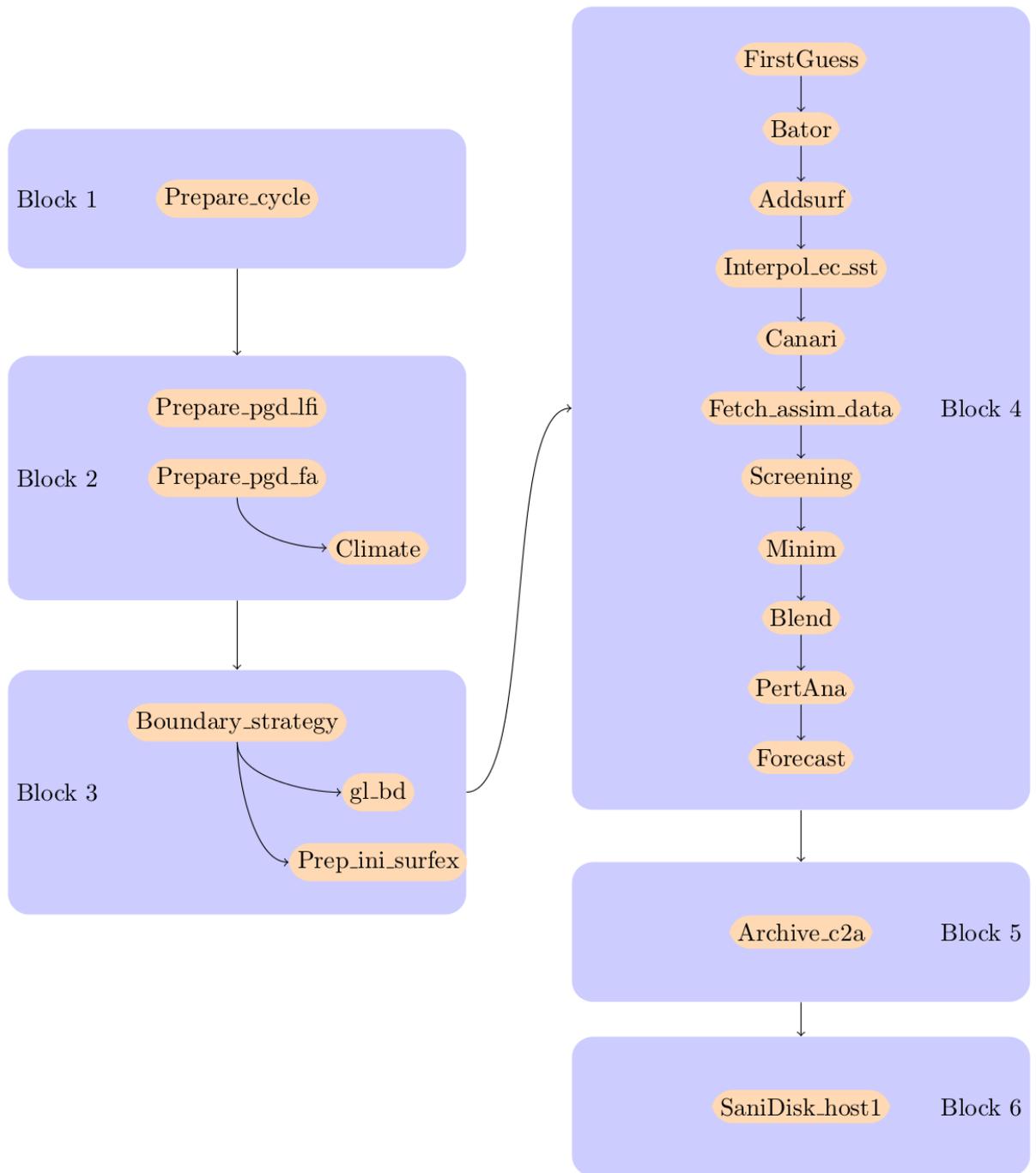

*Figure 9: graphical representation of the dependencies of the profiled RMI-EPS jobs. A black arrow indicates a dependency between jobs/blocks.*

### 5.3.1 Profiling

The scripts of the chosen jobs were adapted such that the appropriate Darshan/IPM modules would be loaded during the execution of the job. Note, however, that darshan and IPM are mutually exclusive! This is because on a Cray system, both tools install their own version of `aprun`. Consequently, two separate runs were required to do all the profiling, one for Darshan and one for IPM.

In the case of Darshan, a distinction was made between the serial parts of a job (which need a `darshan-single` module) and the parallel part (which need a `darshan` module), so within the same run we needed to switch between one and the other at the right times. Correspondingly, two sets of Darshan output files were produced which need to be combined along with the IPM output during the ingestion step (see below).

### 5.3.2 Ingestion

The next step in constructing a synthetic workload model of a given code using the `Kronos` suite consists of ingesting the previously obtained profiling data. This is done with the `kronos-ingest` command which reads one or more profile datafiles from a given job from either `Darshan` or `IPM` and turns them into a `Kronos`-specific `pkl` file.

Subsequently, the `Darshan` and `IPM` `pkl` files of the same job (both the single and parallel ones) need to be combined to produce a complete description of that job. The `Kronos` suite does this with the help of a python script that can be controlled in an interactive way, allowing some additional input/output customization by the user. The end product of this step is a `Kronos`-specific `kpf` file of each individual job.

### 5.3.3 Schedule generation

The final step in the preparation of a synthetic model is the generation of `ksf` files. These files combine the profiling data of individual jobs (i.e. the content of `kpf` files) along with information on how the various jobs depend on each other. The RMI-EPS ensemble suite consists of many individual jobs that are run in succession. A job is submitted only if the jobs that produce its input have completed. The entire ensemble suite therefore consists of a tree of dependencies between jobs. For `Kronos` to be able to model the suite, it needs to know this dependency tree (as shown in Figure 9).

To achieve all of this, the `kpf` files are first run through the tool called `kronos-model`. The result is a `ksf` file of the entire suite containing all the profiling data. Inside this `ksf` file, every individual job is given a job-id. Finally, for every job, the list of other job-id's on which it depends are filled in manually. For `Kronos`, the resulting `ksf` file now represents a complete description of the entire RMI-EPS suite.

### 5.3.4 Execution

The `ksf` file can now be used by a tool called `kronos-executor` to make a simulated run of the RMI-EPS suite. From the description inside the `ksf` file, `kronos`-executor submits synthetic individual jobs to the queuing system of the cluster following the prescribed job dependencies. The interesting part of the `Kronos` suite is that the `ksf` files can be modified to run scaled version of the modeled code. For instance, one can prescribe scenarios involving more intense I/O workloads. This allows to simulate how the code as well as the cluster itself would behave in the case of e.g. higher resolution grids. Typically, the ECMWF needs to be notified when a run of a synthetic model of a complex application like the RMI-EPS suite is performed, since it can disrupt the normal workload on the entire cluster, especially if a scaled-up version of the suite is simulated. The explicit request allows the sys-admins to take the necessary precautions to minimize the burden on other users.

### 5.4 Discussion

At this stage, we have performed a proof of concept test. We constructed a .ksf file which represents the workload of one of the control members of the RMI-EPS ensemble. The file contained the profiling results of the individual jobs listed in Table 1. This workload was small enough to be run on the *lxg* server at ECMWF. The run completed successfully, but due to time constraints, we do not include specific results in this deliverable.

Ultimately, the main result is that we were able to model the RMI-EPS suite using `Kronos` proving that `Kronos` can be a useful tool for the NWP community to assist in the design of benchmarks towards vendors. But in the context of ESCAPE, we will nevertheless also require a tool that can simulate wall-clock time and energy consumption on future hardware.

## 6 Conclusion

In this deliverable a more in-depth description was given of the RMI-EPS ensemble suite. We have provided a detailed report on the workflow of the suite and defined 5 main categories of jobs; pre-processing, LBCs, data assimilation, forecast and post-processing.

Combined Energy and wall-clock time measurements of the entire RMI-EPS suite were performed and they indicate that the wall-clock times are relatively spread between the various defined job categories, with the forecast accounting for the largest fraction at about 35%. As far as energy consumption is concerned, the forecast part dwarfs everything else and is responsible for up to 99% of the total energy consumption. This means that energy optimizations for the forecast part will translate almost proportionally into optimizations of the whole suite, while the maximum theoretical speed-up due to forecast optimizations cannot exceed a factor of about 3/2. Therefore, in terms of energy consumption, optimizations should first focus on the forecast part. For wall-clock time performance gains, however,

ESCAPE 2018optimizations (and possibly additional dwarfs) can be considered for the categories outside of the forecast part.

Finally, we described the necessary steps to build a synthetic model of the suite through the `Kronos` workload simulator. Such a synthetic model allows predicting the I/O and MPI behavior of the suite while subjected to hypothetical workloads on existing hardware. This is meant as a proof of concept and the necessary workflow is described without providing results of actual simulations.

# 7 References

Bonanni, A., Smart S. D., Quintino T. (2016): `Kronos` : benchmarking HPC systems with realistic workloads.
http://www.nextgenio.eu/sites/default/files/documents/publications/A0_portrait_ISC17_NEXTGenIO_Kronos_rev2.pdf

Smet, G. (2017). RMI-EPS: a prototype convection-permitting EPS for Belgium. ALADIN-HIRLAM Newsletter No. 8, 73-79.D4.6 Report on workflow analysis for specific LAM applications    21

ESCAPE 2018

## Document History

| Version | Author(s) | Date | Changes |
|---|---|---|---|
| 0.1 | Joris Van Bever, Geert Smet, Daan Degrauwe | 9/5/2018 | original |
| 1.0 | Joris Van Bever | 28/5/2018 | Internal review corrections/suggestions |
|  |  |  |  |
|  |  |  |  |

## Internal Review History

| Internal Reviewers | Date | Comments |
|---|---|---|
| Peter Messmer (NVIDIA) | 17/05/2018 | Approved with comments |
| Xavier Vigouroux (BULL) | 18/05/2018 | Approved with comments |
|  |  |  |
|  |  |  |

## Effort Contributions per Partner

| Partner | Efforts |
|---|---|
| RMI | 1.5 |
|  |  |
|  |  |
| **Total** | 0 |

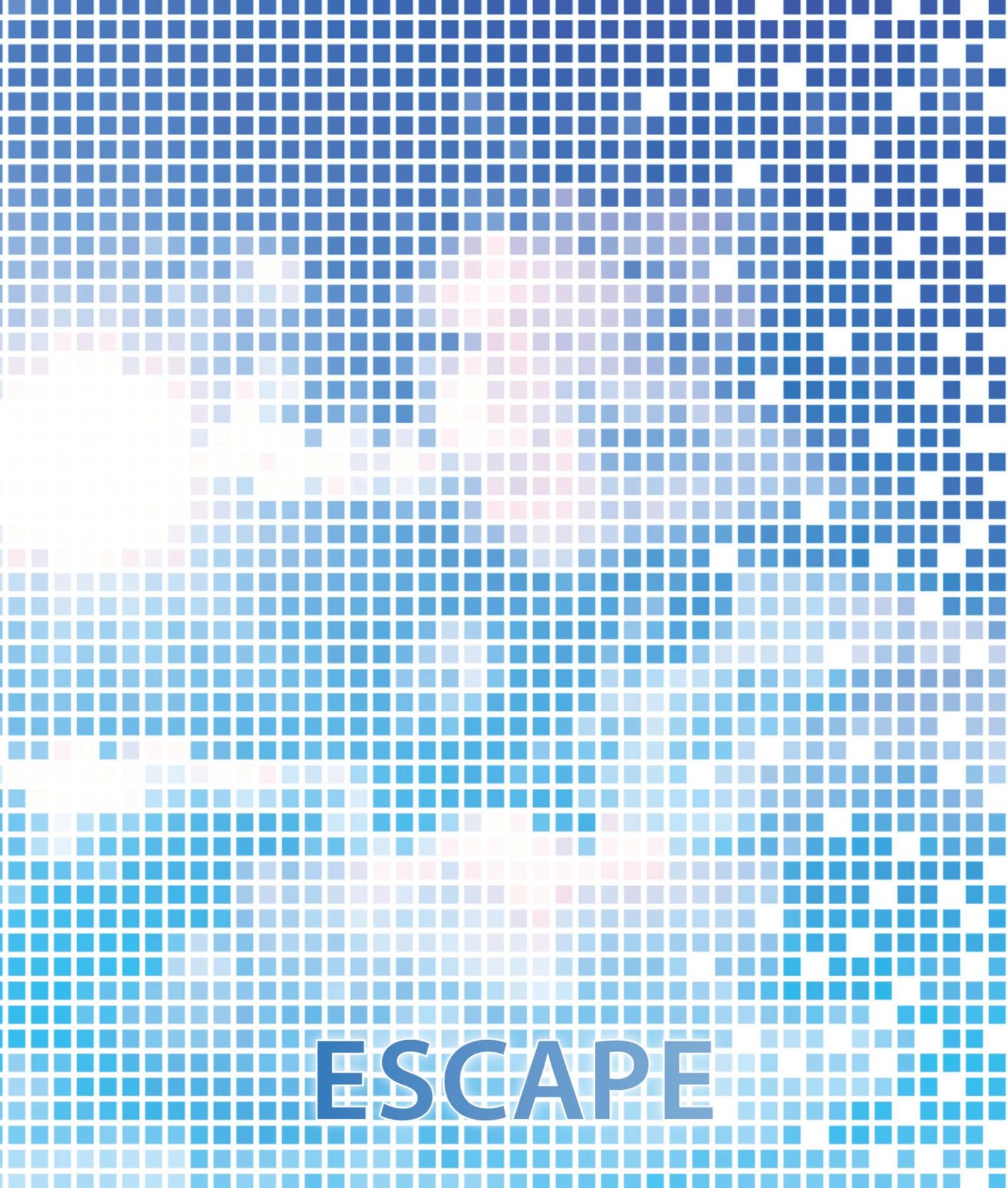

ESCAPE